\documentclass[twocolumn,superscriptaddress,pra,aps,showpacs]{revtex4}
\usepackage{amsfonts}
\usepackage{amsmath}
\usepackage{epsfig}

\begin{document}

\title{Complex Dynamics of Correlated Electrons in Molecular Double Ionization by  an
Ultrashort Intense Laser Pulse}
\author{J. Liu}
\affiliation{Institute of Applied Physics and Computational
Mathematics, P.O.Box 100088, Beijing, P. R. China}
\author{D.F. Ye}
\affiliation{Institute of Applied Physics and Computational
Mathematics, P.O.Box 100088, Beijing, P. R. China}
\author{J. Chen}
\affiliation{Institute of Applied Physics and Computational
Mathematics, P.O.Box 100088, Beijing, P. R. China}
\author{X. Liu}
\affiliation{State Key Laboratory of Magnetic Resonance and Atomic
and Molecular Physics, Wuhan Institute of Physics and Mathematics,
Chinese Academy of Sciences, Wuhan 430071, P. R. China}

\begin{abstract}
With a semiclassical quasi-static model we achieve an insight into
the complex dynamics of two correlated electrons under the combined
influence of a two-center Coulomb potential and an intense laser
field. The model calculation is able to reproduce experimental data
of nitrogen molecules for a wide range of laser intensities from
tunnelling to over-the-barrier regime, and predicts a significant
alignment effect on the ratio of double over single ion yield. The
classical trajectory analysis allows to unveil sub-cycle molecular
double ionization dynamics.
\end{abstract}

 \maketitle

Within the strong-field physics community, there has been increasing
interest on double ionization (DI) of molecules in intense laser
pulses and a large variety of novel phenomena  has emerged. The
diatomic molecules show a much higher double ionization yield than
the prediction of the single-active-electron (SAE) model by many
orders of magnitude \cite{guo,cornaggia}, and DI yield as well as
ionized-electron momentum distribution exhibit a strong dependence
on molecular structure and alignment \cite{alnaser,eremina,zeidler}.
Experimental data indicate that a rescattering mechanism is
responsible for nonsequential double ionization (NSDI) caused by
strong correlation between two electrons. In this process, which has
been extensively investigated for atoms \cite{becker}, an electron
freed by tunnelling ionization is driven by the laser electric field
into a recollision with its parent ion. This essentially classical
rescattering picture implicitly suggests that the well synchronized
recolliding electron burst with respect to the laser field be an
alternative attosecond pulse source for the probe of molecular
dynamics \cite{Niikura}. However the details of this pivotal
recolliding event for molecules remain unknown.

The complexity of the dynamics of the two correlated electrons
responding to a two-center nuclear attraction and the laser force,
on the other hand, poses a great challenge to any theoretical
treatment. For instance, a time-dependent, three-dimensional quantum
mechanical computation from first principles has not yet been
accomplished even for the simpler case of atoms \cite{taylor}. This
leaves approximate approaches developed recently, such as
one-dimensional quantum model \cite{pegarkov}, many-body S-matrix
\cite{beck} and simplified classical methods \cite{saddlepoint}.
However, the complex electron dynamics which is crucial for
molecular DI is still not fully explored and the theoretical results
can not account for experimental data quantitatively. In this
letter, we employ a feasible semiclassical theory, i.e. an \emph{ab
initio} 3D calculation including classical rescattering and quantum
tunnelling effects, providing an intuitive way of understanding the
complex dynamics involved in the molecular DI. Our calculation is
capable of reproducing unusual excess DI rate for a wide range of
laser intensities quantitatively (see Fig. 1), thus consolidating
the classical rescattering view of molecular DI. In particular, with
classical trajectory analysis, we are able to unveil the sub-cycle
dynamics behind molecular DI and predict a significant influence of
molecular alignment on the ratio of double over single ion yield.

% Our
%model also predicts a peak shift of $30^o$ off the field maximum for
%the laser phase at ionization that accounts for the observed
%accumulation of ionized electrons in the first and third quadrants
%of momentum plane\cite{zeidler}. The analysis of correlated electron
%trajectories thus unveils the fully dynamics behind DI of molecule.

%%%%%%%%%%%%%%%%%%%%%%%%%%%%%%%%%%%%%%%%%%%%%%%%%%%%%%%%%%%%%%%%%%%%%%%%%%%%%%%%%%%%%%%%%
\begin{figure}[!b]
\begin{center}
\rotatebox{0}{\resizebox *{7.5cm}{5cm} {\includegraphics
{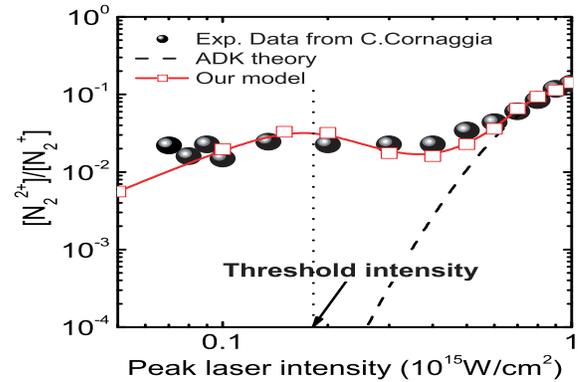}}}
\end{center}
\caption{(color online). Comparison between DI data\cite{cornaggia}
and theory for nitrogen molecule. 0.185 PW/cm$^2$ is the threshold
intensity separates the tunnelling and over-the-barrier regime as
schematically plotted in Fig.2. In the calculation, the laser
frequency $\omega$ of $0.05695a.u.$ and the number of optical cycle
of 37 are chosen to match the experiments of Cornaggia
\cite{cornaggia}. To our knowledge, the results from our
calculations are the first to be in good agreement with experimental
data for a wide range of laser intensities from tunnelling to
over-the-barrier regime.} \label{37T}
\end{figure}
%%%%%%%%%%%%%%%%%%%%%%%%%%%%%%%%%%%%%%%%%%%%%%%%%%%%%%%%%%%%%%%%%%%%%%%%%%%%%%%%%%%%%%%%%%

\begin{figure}[t]
\begin{center}
\rotatebox{0}{\resizebox *{5.0cm}{3.0cm} {\includegraphics
{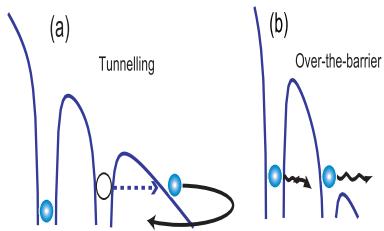}}}
\end{center}
\caption{(color online). Schematic representation of the ionization
mechanisms addressed in this work. (a) Tunnelling ionization. (b)
Over-the-barrier ionization.
%(c) The contour plot of the combined Coulomb potential
%and external field.
%It clearly shows that the saddle point locates
%approximately along the direction of the external field.
}
\label{firstion}
\end{figure}

The model we propose here is in the spirit of that of semiclassical
treatment of DI of atoms in high-intensity field \cite{liuchenfu}.
We consider a molecule composed of two nucleus and two valence
electrons interacting with an infrared laser pulse. When the laser
intensity is smaller than a threshold value (see Fig. \ref{37T}),
one electron is released at the outer edge of the suppressed Coulomb
potential through quantum tunnelling (Fig. \ref{firstion}(a)) with a
rate $\varpi(t_{0})$ given by molecular ADK formula \cite{adk}. The
initial position of the tunnelled electron can be derived from the
equation, $
-\frac{1}{r_{a1}}-\frac{1}{r_{b1}}+\int\frac{\left\vert \Psi(\mathbf{r}%
^{^{\prime}})\right\vert ^{2}}{\left\vert \mathbf{r}_{1}\mathbf{-r}%
^{^{\prime }}\right\vert }d\mathbf{r}^{^{\prime}}+I_{p1}-z_{1}%
\varepsilon(t_{0})=0\, $ with $x_{1}=y_{1}=0$. The wavefunction
$\Psi$ is given by the linear combination of the atomic
orbital-molecular orbital (LCAO-MO) approximation \cite{liuchen}.
%Taking $N_{2}^{+}$ for example, we choose $\phi(r)=\frac{\lambda^{3/2}}{%
%\sqrt{\pi}}e^{-\lambda r}$ as the trial function to construct the
%molecular orbital $\Psi(r)=c[\phi(r_{a2})+\phi(r_{b2})]$, where c is
%the normalization factor. The parameter $\lambda$, which equals to
%1.54 for $N_{2}^{+}$, is determined through variational approach.
The initial velocity of tunnelled electron is set to be $(v_{\perp}\cos\varphi,v_{\perp}\sin%
\varphi,0)$, where $v_{\perp}$ is the quantum-mechanical transverse
velocity distribution satisfying $
w(v_{\perp})dv_{\perp}=\frac{2(2I_{p1})^{1/2}v_{\perp}}{\varepsilon(t_{0})}%
\exp(-\frac{v_{\perp}^{2}(2I_{p1})^{1/2}}{\varepsilon(t_{0})})dv_{\perp
}$, and $\varphi$ is the polar angle of the transverse velocity
uniformly distributed in the interval $[0,2\pi]$ \cite{liuchenfu}.
%%%%%%%%%%%%%%%%%%%%%%%%%%%%%%%%%%%%%%%%%%%%%%%%%%%%%%%%%%%%%%%%%%%%%%%%%%%%%%%%%%%%%%%%%
%%%%%%%%%%%%%%%%%%%%%%%%%%%%%%%%%%%%%%%%%%%%%%%%%%%%%%%%%%%%%%%%%%%%%%%%%%%%%%%%%%%%%%%%%%
For the bound electron, the initial position and momentum are
depicted  by single-electron microcanonical distribution (SMD)
\cite{smd}$, F(\mathbf{r}_{2},\mathbf{p}_{2})=k\delta\lbrack I_{p2}-\mathbf{p}%
_{2}^{2}/2-W(r_{a2},r_{b2})],$ where $k$ is the normalization
factor, $I_{p2}$ denotes the ionization energy
of molecular ions, and $%
W(r_{a2},r_{b2})=-1/r_{a2}-1/r_{b2}$ is the total interaction
potential between the bound electron and two nuclei.

The above scheme is only applicable when the laser intensity is
lower than the threshold value \cite{liuchen}. To give a complete
description of the DI of molecular system for the whole range of the
laser intensities (see Fig. 1), one need to extend the above model
to the over-the-barrier regime (Fig. 2b). This is done by
constructing the initial conditions with double-electron
microcanonical distribution (DMD) \cite{dmd}, i.e.,
$F(\mathbf{r}_{1},\mathbf{r}_{2},\mathbf{p}_{1},\mathbf{p}_{2})=%
\frac{1}{2}[f_{\alpha }(\mathbf{r}_{1}\mathbf{,p}_{1})f_{\beta }(\mathbf{r}%
_{2}\mathbf{,p}_{2})+f_{\beta }(\mathbf{r}_{1}\mathbf{,p}_{1})f_{\alpha }(%
\mathbf{r}_{2}\mathbf{,p}_{2})],$ with
$f_{\alpha ,\beta }(\mathbf{r,p})=k\delta \lbrack I_{p1}-\frac{\mathbf{p}^{2}%
}{2}-W(r_{a},r_{b})-V_{\alpha ,\beta }(\mathbf{r)}], $ where $
V_{\alpha ,\beta }(\mathbf{r)=}\frac{1}{r_{b,a}}[1-(1+\kappa
r_{b,a})e^{-2\kappa r_{b,a}}] $ represents the mean interaction
between the electrons, $\kappa $ can be obtained by a variational
calculation of the ionization energy of molecules.

The subsequent evolution of the two-electron system with the above
initial conditions is simulated by the classical Newtonian equations
of motion: $
%\begin{equation}
\frac{d^{2}\mathbf{r}_{i}}{dt^{2}}=\mathbf{\varepsilon }(t)-%
\bigtriangledown(V_{ne}^{i}+V_{ee}).
%\label{Newton}
%\end{equation}
$ Here index i denotes the two different electrons, $V_{ne}^{i}$ and
$V_{ee}$ are Coulomb interaction between nuclei and electrons and
between two electrons, respectively. $V_{ne}^{i}
=-\frac{1}{r_{ai}}-\frac{1}{r_{bi}}, V_{ee} =
\frac{1}{\left\vert \mathbf{r}_{1}-\mathbf{r}_{2}\right\vert },$ where $%
r_{ai}$ and $r_{bi}$ are distances between the ith electron and
nucleus a and b. The above Newtonian equations of motion are solved
using the 4-5th Runge-Kutta algorithm and DI events are identified
by  energy criterion. In our calculations, more than $10^5$ weighted
(i.e., by rate $\varpi(t_{0})$) classical two-electron trajectories
are traced and a few thousands or more of DI events are collected
for statistics. Convergency of the numerical results is further
tested by increasing the number of launched trajectories twice.

The above model is applied to the study on DI of molecular nitrogen.
We first calculate the ratio between the double and single ionization yield with respect to the peak laser intensities from $%
5\times 10^{13}$W/cm$^{2}$ to $1\times 10^{15}$W/cm$^{2}$. In Fig.
\ref{37T} the calculated results is compared with that of recent
experiments of Cornaggia \emph{et.al.} \cite{cornaggia} and a good
agreement is obtained for such a broad range of laser intensities.

\begin{figure}[t]
\begin{center}
\rotatebox{0}{\resizebox *{8.0cm}{8.5cm} {\includegraphics
{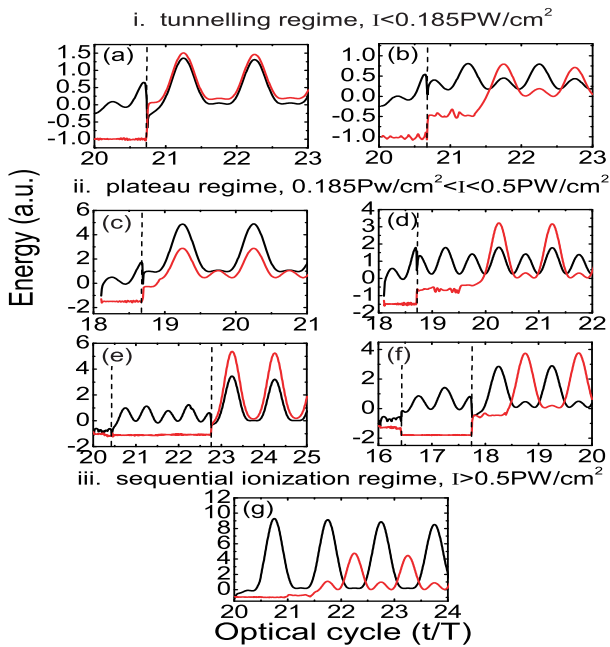}}}
\end{center}
\caption{(color online). Typical energy evolution of the electron
pair in different laser intensity regime. Vertical dashed lines
indicate the moment when collision between electrons emerge.}
\label{evl}
\end{figure}

In the following, we proceed to explore the correlated electron
dynamics responsible for DI of molecular nitrogen. The classical
trajectory method allows us to select out the individual DI
trajectories and back analyze their dynamics in detail. The typical
electron trajectories are shown in Fig. 3, presented in an energy
versus time plot. The threshold value of 0.185 PW/cm$^2$ separates
the DI data into two parts. When the peak laser intensity is below
this value, there exist two dominant processes responsible for
emitting both electrons, namely, collision-ionization(CI) and
collision-excitation-ionization(CEI), as shown in Fig.
\ref{evl}(a)(b), respectively. For CI, the tunnelled electron is
driven back by the oscillating laser field to collide with the bound
electron near its parent ion causing an instant ($\sim$ attosecond)
ionization. For CEI, DI event is created by recollision with
electron impact excitation followed by a time-delayed ($\sim$ a few
optical periods) field ionization of the excited state. When the
laser intensity is above the threshold value, over-the-barrier
ionization emerges. In this regime we observe more complicated
trajectories for DI processes. Except for CI (Fig. 3(c)) and CEI
(Fig. 3(d)) trajectories similar to tunnelling case, there are
multiple-collision trajectories as shown in Fig. 3(e),(f) as well as
collisionless trajectory of Fig. 3(g). In Fig. 3(e) and (f),
initially two valence electrons entangle each other, experience a
multiple-collision and then emit. The four types of trajectories
shown in Fig. 3(c-f) represent the dominant processes of DI in the
plateau regime from 0.185PW/cm$^2$ to 0.5PW/cm$^2$, each of them
accompanied by one or multiple times of collisions between two
electrons \cite{paulus, ho, liux}. However, above 0.5PW/cm$^2$, DI
is dominated by a collisionless sequential ionization whose typical
trajectory is represented by Fig. 3(g). In this regime results from
our model agree with ADK theory.
%%%%%%%%%%%%%%%%%%%%%%%%%%%%%%%%%%%%%%%%%%%%%%%%%%%%%%%%%%%%%%%%%%%%%%%%%%%%%%%%%%%%%%%%%
%%%%%%%%%%%%%%%%%%%%%%%%%%%%%%%%%%%%%%%%%%%%%%%%%%%%%%%%%%%%%%%%%%%%%%%%%%%%%%%%%%%%%%%%%%

%%%%%%%%%%%%%%%%%%%%%%%%%%%%%%%%%%%%%%%%%%%%%%%%%%%%%%%%%%%%%%%%%%%%%%%%%%%%%%%%%%%%%%%%%
\begin{figure}[t]
\begin{center}
\rotatebox{0}{\resizebox *{8.0cm}{7cm} {\includegraphics
{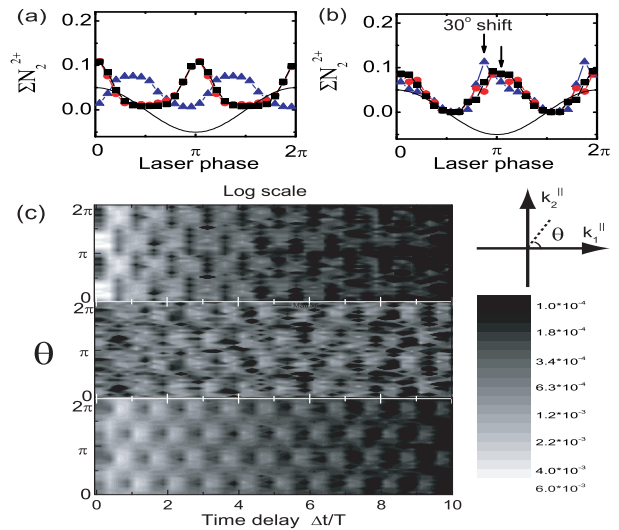}}}
\end{center}
\caption{(color online). DI yield vs laser phase when (a) the two
electrons become closest; (b) both the electrons are ionized, at
different laser intensity $0.12$PW/cm$^{2}$(triangle),
$0.4$PW/cm$^{2}$(circle) and $1$PW/cm$^{2}$(square), respectively.
(c) The relationship between the correlated momentum and the delay
time at $0.12$PW/cm$^{2}$ (upper), $0.4$PW/cm$^{2}$ (mid) and
$1.0$PW/cm$^{2}$ (low), respectively. } \label{timeall}
\end{figure}
%%%%%%%%%%%%%%%%%%%%%%%%%%%%%%%%%%%%%%%%%%%%%%%%%%%%%%%%%%%%%%%%%%%%%%%%%%%%%%%%%%%%%%%%%%
The analysis of trajectories of electron-electron pairs may
provide insight into the complicated dynamics of DI with sub-cycle
time resolution, and the important information is revealed by the
laser field phase at the moments of collision and ionization
\cite{feuerstein, weckenbrock}. We choose three typical laser
intensities, $0.12$PW/cm$^{2}$, $0.4$PW/cm$^{2}$ and
$1$PW/cm$^{2}$, representing the tunnelling, plateau and
sequential ionization regime, respectively.

Fig. \ref{timeall}(a) shows the diagram of DI yield versus laser
phase at the moment of closest collision. In the tunnelling regime
(i.e., $0.12$PW/cm$^{2}$), we note that the collision can occur
throughout most of the laser cycle and the peak emerges slightly
before the zeroes of the laser field. This is consistent with the
prediction of simple-man model \cite{corkum} and recent results from
purely classical calculation \cite{haan}. However, for the other two
cases, the collision between the two correlated electrons occurs
mainly at peak laser field. This is because the ionization mechanism
changes at the transition to over-the-barrier regime, where both
electrons rotate around the nuclei and their distance could be very
close before one of them is driven away by the external field.

Fig. \ref{timeall}(b) plots DI yield as a function of the laser
phase at the instant of ionization. Most DI occurs around the
maximum of laser field for different intensity regime.
Interestingly, for the tunnelling case, we observe a peak shift of
$\sim 30^o$ off the field maximum. With assuming that the colliding
electron leaves the atom with no significant energy and
electron-electron momentum exchange in final state is negligible
\cite{assuming}, the parallel momentum $k_{1,2,}^{||}$ of each
electron results exclusively from the acceleration in the optical
field: $k_{1,2}^{||}=\pm 2\sqrt U_p \sin\omega t_{ion}$
\cite{weckenbrock}. The above shifted peak indicates the
accumulation of the emitted electrons at $k_{1}^{||}=k_{2}^{||}=\pm
\sqrt U_p $ in the first and third quadrants of parallel momentum
plane $(k_{1}^{||}, k_{2}^{||})$. It is  consistent with the
experimental data of Ref. \cite{zeidler} (see their Fig. 2).

Fig. \ref{timeall}(c) shows the phase angle of momentum vector
$(k_{1}^{||},k_{2}^{||})$ with respect to the  delayed time between
the closest collision  and ionization. The integration over the
phase angle gives total DI yield versus the delayed time. In all
three cases we observe a long-tail up to several optical periods.
For the sequential ionization of 1PW/cm$^2$, it means that the
second electron is slowly (i.e., waiting for a few  optical cycles)
ionized after the first electron is deprived from nuclei by the
laser field. In the tunnelling regime, the long-tail indicates that
CEI mechanism is very pronounced for the molecular DI and
contributes to $\sim~$80\% of the total DI yield.

This observation is different from purely classical simulation
\cite{haan}, where CI effect is believed to be overestimated. Our
results, however, are consistent with experimental data for Ar atom
\cite{feuerstein}, where ionization potential and laser field
parameters are close to our case. The reason is stated as follows:
For the intensity of $0.12$PW/cm$^{2}$, the maximal kinetic energy
of the returned electron is $3.17U_{p}=0.85a.u.$, still smaller than
the ionization energy of $N_{2}^+$. Even with the assistance of the
Coulomb focusing \cite{brabec}, it is not easy for the returned
electrons to induce too many CI events. Furthermore, such time delay
might provide more physics beyond simple rescattering scenario.
Recently a statistical thermalization model has been proposed for
the nonsequential multiple ionization of atoms in the tunneling
regime \cite{attothermal}. This model shows that sharing of excess
energy between the tunnelled electron and the bound electrons takes
some time, resulting in a time delay on attosecond time scale
between recollision and ionization. Our simulation upholds this
picture of attosecond electron thermalization: on upper panel of
Fig. \ref{timeall}(c), two bright spots are observed at a similar
time delay on subfemtosecond time scale for CI trajectory.

The regular patterns in upper and lower panels of Fig.
\ref{timeall}(c) exhibit that the ejection of electrons in the
same-hemisphere and opposite-hemisphere emerge alternately with
respect to the delayed time. For a time delay of odd
half-laser-cycles, two electrons emit in the same direction. In
contrast, they emit in the opposite direction for an even
half-laser-cycles time delay.
%The above observation supports the picture of
%attosecond electron thermalization of tunnelling reigme
%\cite{attothermal}.
In mid of Fig. \ref{timeall}(c), on the other hand, the irregular
pattern emerges as the signature of complicated multiple-collision
trajectories for DI in the plateau regime. It implies that the
trajectories of two electrons entangle with each other before DI
ionization occurs and the electrons' motion might be chaotic
\cite{liuhu}.

%%%%%%%%%%%%%%%%%%%%%%%%%%%%%%%%%%%%%%%%%%%%%%%%%%%%%%%%%%%%%%%%%%%%%%%%%%%%%%%%%%%%%%%%%
\begin{figure}[t]
\begin{center}
\rotatebox{0}{\resizebox *{7.5cm}{4cm} {\includegraphics
{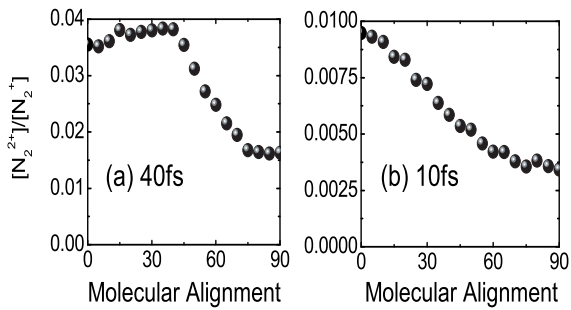}}}
\end{center}
\caption{(color online). The molecular alignment dependence of DI
ratios for laser intensity of  0.15PW/cm$^{2}$. } \label{angle}
\end{figure}
%%%%%%%%%%%%%%%%%%%%%%%%%%%%%%%%%%%%%%%%%%%%%%%%%%%%%%%%%%%%%%%%%%%%%%%%%%%%%%%%%%%%%%%%%%
When the light intensity is high enough, it has been consensus that
DI behavior of atoms is determined by essentially electron physics
in the presence of laser field \cite{haan,liuchenfu}. Good
correspondence between our theoretical calculations and experimental
data confirms the validity of the above picture in molecular DI
case. In our model, after tunnelling electrons travel much of the
time in the intense laser field like a classical object and solely
electron collision physics determines the fate of DI of molecules.
However, the inherent nuclear degree of freedom of molecule do
manifest themselves as the significant alignment effect in our
model. To clearly demonstrate it, we calculate the ratios between
double and single ionization at different molecular alignment
angles. Main results are presented in Fig. \ref{angle}. It shows
that, i) The ratio between  DI  and single-ionization yield is less
for perpendicular molecules than that of parallel molecules; ii)
This anisotropy becomes more dramatic for a shorter laser pulse.
Further explorations show that molecular alignment also
significantly affects the correlated momentum distribution of
emitted electrons. Details will be presented elsewhere \cite{li}.

In summary, we exploit a semiclassical quasi-static model to achieve
insight into the correlated electron dynamics in molecular DI under
the relevant experimental conditions, i.e., highly nonperturbative
fields with femtosecond or shorter time resolution. Our calculation
unveils sub-cycle dynamics behind molecular DI and predicts a
significant influence of the molecular alignment on the ratio of
double over single ion yield. Because molecular alignment is
controllable with present technique \cite{zeidler} the above results
can be regarded as our theoretical prediction which may be tested in
future experiments.

This work is supported by NNSF of China No.10574019,  CAEP
Foundation 2006Z0202, and 973 research Project No. 2006CB806000.
We thank J. H. Eberly stimulating discussions and are indebted to
C. Figueira de Morisson Faria for reading the manuscript carefully
and useful suggestions.


\begin{thebibliography}{}
\bibitem{assuming} Note that these assumptions have been
checked by directly tracing the trajectories.
\end{thebibliography}
\end{document}